\documentstyle[epsfig]{aipproc}
\begin{document}
%\input epsf
%  laur 97-5163  LAUR LA UR LA-UR
\title{Variability in Shell Models of GRBs}
\author{M. C. Sumner and E. E. Fenimore}
\address{NIS-2, MS D436, Los Alamos National Laboratory, Los Alamos, NM
        87545}
\maketitle
\begin{abstract}
  Many cosmological models of gamma-ray bursts (GRBs) assume that a
  single relativistic shell carries kinetic energy away from the source
  and later converts it into gamma rays, perhaps by interactions with
  the interstellar medium or by shocks within the shell.
  Although such models are able to reproduce general trends in GRB time
  histories, it is difficult to reproduce the high degree of variability
  often seen in GRBs.  We investigate methods of achieving this
  variability using a simplified external shock model.  Since our model
  emphasizes geometric and statistical considerations, rather than the
  detailed physics of the shell, it is applicable to any theory that
  relies on relativistic shells.  We find that the variability in GRBs
  gives strong clues to the efficiency with which the shell converts its
  kinetic energy into gamma rays.
\end{abstract}
The ``external shock'' models of gamma-ray bursts (GRBs) assume that the
gamma rays are emitted from a single shell of material traveling at
highly relativistic speeds (for example, \cite{msef:Meszaros} and
\cite{msef:Piran}).  Fenimore, Madras, and Nayakshin \cite{msef:Madras}
have found that the envelope of emission for a single relativistic shell
fits the overall shape of a few GRBs, but it does not account for the
wide variability found in GRBs.  In this paper, we
investigate methods of achieving the observed degree of variability in
GRB time histories using randomly placed active regions on the shell.

\section*{Description of the Model}
The ``time'' that is measured in
GRB time histories is the time at which photons arrive at the
detector (denoted by $T$), and is
not the time of emission as would be measured in the rest frame of the
detector (equivalent to the rest frame of the  shell's expansion,
denoted by $t$).
 We set $t=0$ to be
the time at which the central explosion occurs, and $T=0$ to be the arrival
time of a photon emitted at $t=0$; therefore
\begin{equation}
T=(1-\beta \cos \theta )t=\Lambda \gamma^{-1}t  
\label{msef:arrival}
\end{equation}
where $\Lambda$ is defined as $\Lambda =\gamma (1-\beta \mu )$, and $\mu
= \cos \theta $. By setting Equation \ref{msef:arrival} equal to a
constant, one finds that the surface of constant arrival time is an
ellipsoid \cite{msef:Rees}. All of the photons emitted from such an
ellipsoid will arrive at the detector simultaneously.

Our model considers a thin shell (as required by the observations
\cite{msef:Madras}) that expands outward from a central point so that
$R=\beta ct$. The shell expands to a radius of $R_{{\rm o}}=\beta
ct_{{\rm o}}$, and then emits gamma rays until it reaches a radius of
$R_{ {\rm max}}=\beta ct_{{\rm max}}$.
The gamma ray emission must
occur from small, independent patches on the shell, which we call
``entities''\cite{msef:Madras}.  Each entity begins as a small
perturbation, perhaps
caused by the shell's interactions with the interstellar medium or by
instabilities within the shell. The entity grows at the sound speed
$c_{s}\sim c$ until it reaches a maximum size of $\Delta R_{\perp
  }=\Gamma c\Delta T_{{\rm p}}$ (see Table 2 in \cite{msef:Madras}).
Thus, each entity represents a causally connected region, and many
entities can fit on the surface of the shell.  We assume that each
entity emits isotropically with a power-law photon spectrum in its own
reference frame, $\Phi ^{^{\prime }}(E^{^{\prime }})=E^{^{\prime
    }}{}^{-\alpha }$ photons (entity $ dt^{^{\prime }}$ $dE^{^{\prime
    }}$ $d\Omega ^{^{\prime }}$)$^{-1}$ (primed quantities are measured
in the rest frame of the entity).
\begin{figure}[tb]
\centering
%\psbox[xsize=0.7\hsize,rotate=n]{2d_3dellipse.eps}
% \epsfig{width=0.7\hsize,file=2d_3dellipse.eps}
\epsfig{width=0.7\hsize,file=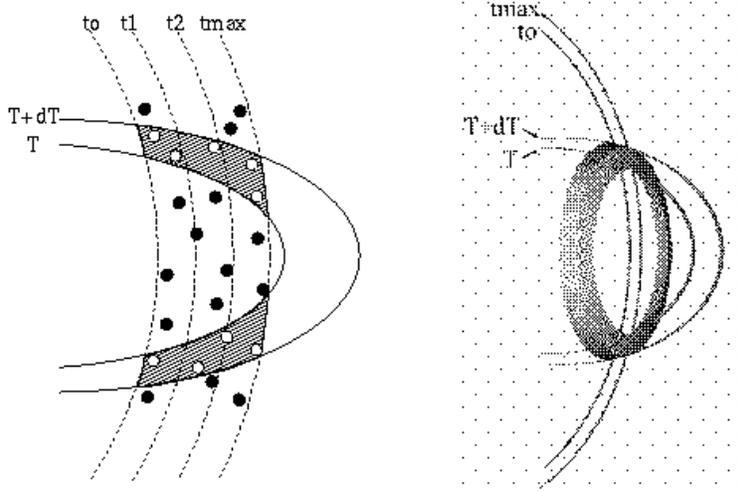}
%\leavevmode \epsfxsize=0.7\hsize \epsfbox{2d_3dellipse.eps}
\caption{{\em a. (left)} The position of the spherical shell at
  four different times as it expands from $t_{{\rm o}}$ to $t_{{\rm
      max}}$.  The two ellipses represent surfaces of constant arrival
  time. Photons emitted from any point on the inner ellipsoid will
  arrive at the detector simultaneously at time $T$, and photons emitted
  from any point on the outer ellipsoid will arrive at the detector at time
  $T+dT$. The cross-hatched region represents the volume that will be
  ``seen'' by the detector between these two times. Each small circle
  represents the volume swept out by an entity during its emission
  lifetime of $\Delta T_{\rm p}$. Only the white entities will be seen
  between $T$ and $T+\Delta T$. {\em b. (right)} A three-dimensional view
  of the cross-hatched area shown at left.  The volume of this annulus
  is given by $\Upsilon(T)dT$.}
\label{msef:2dellipse}
\label{msef:3dellipse}
\end{figure}

Figure \ref{msef:2dellipse} gives a pictorial representation of our
model.  It is important to emphasize that we are considering a {\em
  thin} shell. Thus, the four dashed lines in Figure \ref
{msef:2dellipse}a represent the shell at four different times. The area
between $t_{{\rm o}}$ and $t_{{\rm max}}$ represents the volume {\em
  swept up} by the shell over its emission lifetime. Likewise, the
entities are small {\em areas} on the shell. The small spheres in Figure
\ref{msef:2dellipse}a represent the volumes swept out by the entities as
the shell expands.

A more quantitative description can be made by calculating the volume
enclosed in the cross-hatched region, which we denote as $\Upsilon
(T)dT$ (Figure \ref{msef:3dellipse}b): 
\begin{equation}
\Upsilon (T)dT=2\pi \int_{\min (R_{{\rm o}},R_{{\rm ell,in)}}}^{\min (R_{%
{\rm max}},R_{{\rm ell,out}})}\int_{\mu (T)}^{\mu (T+dT)}r^{2}d\mu \ dr\ 
\end{equation}
where $R_{{\rm ell,out}}=\beta c(T+dT)/(1-\beta )$, and $R_{{\rm
    ell,in}}=\beta cT/(1+\beta )$. Equation \ref{msef:arrival} gives
$\mu (T)=\left[ \beta ^{-1}-Tc/r\right] $, and $\mu (T+dT)=\left[ \beta
  ^{-1}-(T+\Delta T)c/r\right] $. Evaluating the integral yields the
volume seen per $dT$:
\begin{eqnarray}
\Upsilon (T) &=&0\hspace{313pt}{\rm if}\ T<T_{{\rm o}}  \nonumber \\
&=&\frac{\pi c\left( \beta c\right) ^{2}}{\left( 1-\beta \right) ^{2}}\left(
T^{2}-T_{{\rm o}}^{2}\right) \hspace*{2.5in}{\rm if}\ T_{{\rm o}}<T<T_{{\rm %
max}}  \nonumber \\
&=&\frac{\pi c\left( \beta c\right) ^{2}}{\left( 1-\beta \right) ^{2}}\left(
T_{{\rm max}}^{2}-T_{{\rm o}}^{2}\right) \hspace{115pt}{\rm if\ }T_{{\rm max}%
}<T<\Gamma ^{2}(1+\beta )^{2}T_{{\rm o}}  \label{msef:volume} \\
&=&\frac{\pi c\left( \beta c\right) ^{2}}{\left( 1-\beta \right) ^{2}}\left(
T_{{\rm max}}^{2}-\frac{T^{2}}{\Gamma ^{4}(1+\beta )^{4}}\right) \hspace{20pt%
}{\rm if}\ \Gamma ^{2}(1+\beta )^{2}T_{{\rm o}}<T<\Gamma ^{2}(1+\beta
)^{2}T_{{\rm max}}  \nonumber \\
&=&0\hspace{253pt}{\rm if}\ T>\Gamma ^{2}(1+\beta )^{2}T_{{\rm max}} 
\nonumber
\end{eqnarray}
where $T_{{\rm o}}=t_{{\rm o}}/(1-\beta )$, and $T_{{\rm max}}=t_{{\rm max}%
}/(1-\beta )$. Note the rather surprising result that the volume ``seen'' by
the detector is constant for $T_{{\rm max}}<T<\Gamma ^{2}(1+\beta )^{2}T_{%
  {\rm o}}$ for the relativistically expanding shell. During these 
times, the number of entities in the cross-hatched region in Figure
\ref{msef:2dellipse}a is a constant with respect to $T$. In comparison,
in the non-relativistic case, the volume seen, and therefore the number
of entities seen, would increase as the area of the shell increased.

\begin{figure}[tb]
\centering
%\psbox[xsize=0.4\hsize,rotate=n]{volumebw.epsi}
\epsfig{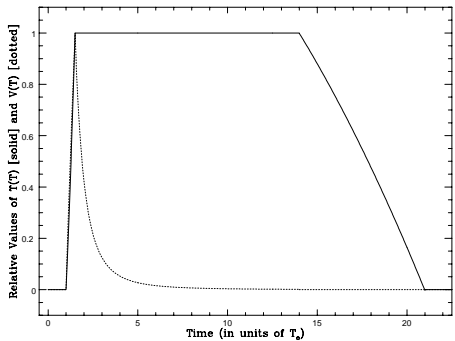}
%\leavevmode \epsfxsize=0.4\hsize \epsfbox{volumebw.epsi}
\caption{The volume seen by the detector per $\Delta T$
  is shown as the solid line, and the resulting photon flux at the
  detector is shown as the dashed line. The volume, $\Upsilon(T)$, and
  the detector signal, $V(T)$, both increase as $T^2$ for $T_{{\rm
      o}}<T<T_{{\rm max}}$; this is identical to the results for a
  non-relativistic shell. However, in the case of a relativistic shell
  for $T > T_{{\rm max}}$, the volume remains constant over a large
  range of $T$, and the signal shows a power-law decay. Since the
  detector sees a constant volume, and thus a constant number of
  entities, per $\Delta T$ for $T>T_{{\rm max}}$, the decrease in the
  photon flux after $T_{{\rm max}}$ is entirely determined by
  relativistic effects.}
\label{msef:volumeplot}
\end{figure}

The time history from such a shell is given by
\begin{equation}
V(T)dT=2\pi \int_{min(R_{{\rm o}},R_{{\rm ell,in)}}}^{min(R_{{\rm max}},R_{%
{\rm ell,out}})}\int_{\mu (T)}^{\mu (T+dT)}\rho C^{\prime }\Lambda
^{-(\alpha +1)}r^{2}d\mu \ dr\ 
\end{equation}
where $V(T)$ is the expected time history in photons $(dT~dA_{det})^{-1}$, $%
C^{^{\prime }}$ is the photon flux as observed in the rest frame of the
entity, and $\Lambda ^{-(\alpha +1)}$ incorporates the relativistic
effects.  We define $\rho$ as the ``density'' of entities so that $\rho~dV$
gives the number of entities within the cross-hatched region in
Figure \ref{msef:2dellipse}a.  (The density $\rho$ is proportional to
the fraction of the shell's surface that emits gamma rays during the
shell's evolution, and therefore gives some idea of how efficiently the
shell converts its kinetic energy into gamma rays.)  The envelope is
\cite{msef:Riken}
\begin{eqnarray}
V(T) &=&0\hspace{300pt}{\rm if}\ T<T_{{\rm o}}  \nonumber \\
&=&\psi \frac{T^{\alpha +3}-T_{{\rm o}}^{\alpha +3}}{T^{\alpha +1}}\hspace{%
200pt}{\rm if}\ T_{{\rm o}}<T<T_{{\rm max}}  \label{msef:envelope} \\
&=&\psi \frac{T_{{\rm max}}^{\alpha +3}-T_{{\rm o}}^{\alpha +3}}{T^{\alpha
+1}}\hspace{147pt}{\rm if}\ T_{{\rm max}}<T<\Gamma ^{2}(1+\beta )^{2}T_{{\rm %
o}}  \nonumber
\end{eqnarray}
where $\psi $ is a constant (see Figure \ref{msef:volumeplot}).

\begin{figure}[tb]
\centering
\epsfig{width=0.8\hsize,file=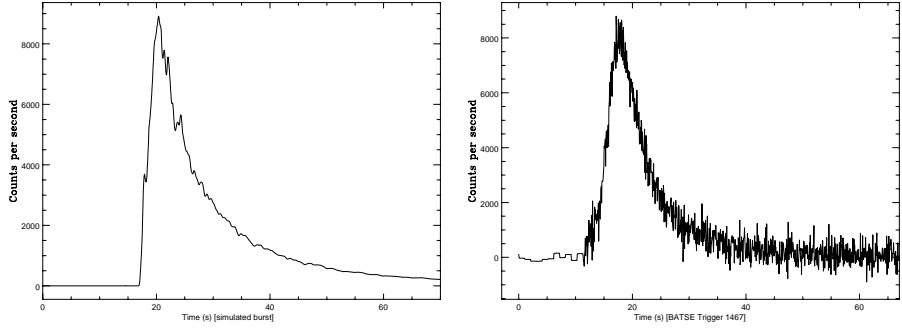}
%\psbox[xsize=0.8\hsize,rotate=n]{d9937_1467.horiz.eps}
%\leavevmode \epsfxsize=0.8\hsize \epsfbox{d9937_1467.horiz.eps}
\caption{{\em a. (left)} Simulated time history using a high density of
  entities.  Nearly 100\% of the shell's surface emitted gamma rays in
  this simulation. Note that the high efficiency gives a smooth profile.
  {\em b. (right)} BATSE trigger 1467.}
\label{msef:678sim}
\end{figure}

\begin{figure}[tb]
\centering
%\psbox[xsize=0.8\hsize,rotate=n]{e9995_678.horiz.eps}
\epsfig{width=0.8\hsize,file=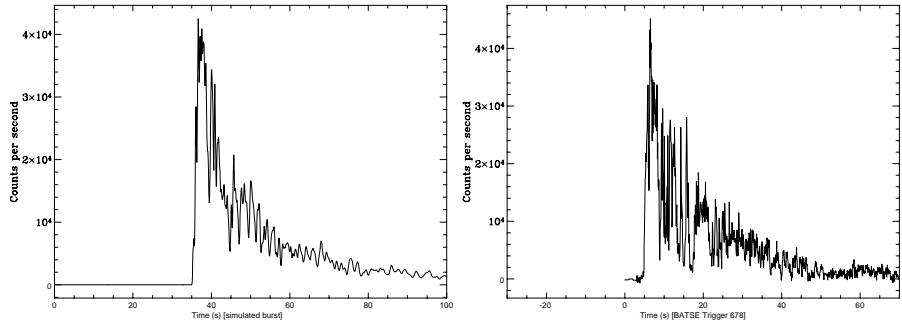}
%\leavevmode \epsfxsize=0.8\hsize \epsfbox{e9995_678.horiz.eps}
\caption{{\em a. (left)} Simulated time history assuming a low density of
  entities.  Only 1\% of the shell's surface emitted gamma rays in this
  simulation.  Note that this low efficiency gives a spiky profile, with
  many ``peaks''. {\em b. (right)} BATSE trigger 678.}
\label{msef:1467sim}
\end{figure}

Equation \ref{msef:envelope} gives the {\em average} signal that would
be expected from a collection of entities scattered over the surface of
a relativistic shell. However, the number of entities seen in any volume
$ \Upsilon (T)dT$ is a random quantity; since the entities are
independent and discrete, the randomness follows simple Poisson
statistics.  Therefore, this model predicts time histories with a mean
given by Equation \ref {msef:envelope} and Poisson variations about that
mean, which would look very similar to the ``peaks'' usually observed in
GRB time histories.  A low value for $\rho $ gives a low number of
entities and correspondingly large Poisson fluctuations, leading to a
``spiky'' time history with many peaks.  Conversely, a high efficiency
leads to a smooth time history.  To illustrate these points, we present
two simulated time histories in Figures \ref{msef:678sim}a and
\ref{msef:1467sim}a with characteristics selected to match particular BATSE
bursts. We estimate that the typical fraction of
a single relativistic  shell's surface that emits gamma rays during
its lifetime is $\sim
10^{-3}$ \cite{msef:Huntsville}.

\section*{Conclusions}
In order to achieve the observed variability in GRBs using a
single-shell model, we have found that the gamma-ray emission must occur
from small patches on the shell (entities).  We have derived two
significant characteristics of the time histories expected from such a
model.
% \begin{enumerate}
% \item 1.
First, the average number of entities contributing to the signal
remains constant throughout much of the time history, although the
overall photon flux decreases due to relativistic effects.
%  \item 2.
Second, the ``peaks'' in a time history can be ascribed to Poisson
variations in the actual number of entities contributing to the signal
at any given time.
% \end{enumerate}
Taken together, these properties imply that the relative variations in a
GRB (i.e. the heights of the peaks relative to the average envelope of
the signal) should remain constant throughout a time history.
Qualitatively, this result is consistent with visual inspections of
GRBs.  In general, bursts  are equally "spiky" during the
first half of the time history as  the second half.

\end{document}